\documentclass[journal=jacsat,manuscript=article]{achemso}

\usepackage{soul}
\usepackage{graphicx}
\usepackage{dcolumn}
\usepackage{bm}

\usepackage[usenames,dvipsnames]{color}
\usepackage{graphicx, microtype}
\usepackage[bookmarks=false,colorlinks]{hyperref}
\hypersetup{linkcolor=Plum,citecolor=RubineRed,filecolor=Plum,urlcolor=PineGreen}

\usepackage{float}
\usepackage{graphicx}
\usepackage{xspace}
\usepackage{ulem}

\usepackage[usenames,dvipsnames]{color}
\usepackage{graphicx,microtype}

\usepackage[bookmarks=false,colorlinks]{hyperref}
\hypersetup{
    linkcolor=Plum,        
    citecolor=RubineRed,             
    filecolor=Plum,      	    
    urlcolor=PineGreen,         
}

\author{Alexandru B. Georgescu}
\email{georgesc@iu.edu}
\affiliation{%
Department of Chemistry, 800 East Kirkwood Avenue, Indiana University, Bloomington, Indiana 47405, United States}
\title{Why Charge Added Using Transition Metals To Some Insulators - Including LK-99 - Localizes and Does Not Yield a Metal}

\begin{document}

\maketitle

\date{\today}%

\begin{abstract}
While adding charge to semiconductors via dopants is a well-established method for tuning electronic properties, we demonstrate that introducing transition metal impurities into certain insulators can lead to localized charge, assisted by a Jahn-Teller distortion. This leads to isolated charge, and an insulating material as opposed to a emergent states - including  superconductivity. We focus on Cu impurities added to Pb$_{10}$(PO$_4$)$_6$O ('LK-99'), replacing 10\% of Cu ions, as discussed in recent literature. Our calculations show that the material remains a wide bandgap insulator with isolated, S=1/2 localized charges on the Cu ions—similar to color centers—even within standard DFT, without the need for electron correlation corrections to the Cu d-orbitals. Superconductivity is excluded by known mechanisms that require the material to be metallic. We resolve previously observed inconsistencies between density functional theory results and experimental findings related to doping site energetics, crystal structure, and transparency. We find that Cu doping either Pb site leads to CuO$_4$ coordination and a similar unit cell volume contraction. Engineering materials with dopant sites that have different local symmetries can induce non-relativistic spin splitting—often referred to as altermagnetism. However, in the case of localized charges, this may enable spins to be individually controlled. 
\end{abstract}

\section{Introduction}

Correlated electronic materials, often materials with partially filled transition metal sites, exhibit a wide range of technologically relevant properties, including magnetism \citep{Tokura1998, Gibertini2019, DivineAlexNPJ, Molegraaf2009, Mizuguchi2008}, metal-insulator transitions \citep{Landscapes2,guzman,Shamblin2018,JMReview,Forst2015,Medarde1998, Claribel,VO2Length,Caviglia2013,Lauren2019,Lucia2023}, and high-temperature superconductivity \citep{FisherEnergy,Lee2006,Norman2020,Gariglio2009,Keimer2015}. The study of impurities, defects, and dopants has a long-standing tradition across various fields, most recently including color centers in diamond for quantum information applications \citep{Hepp2014,JESENOVEC2022126419}, and the use of transition metal centers at solid surfaces in catalysis \citep{Chakraborty2022}. 

Underlying correlated electron phenomena is the interplay between local electronic states (often d or f states, flatbands \citep{Yin2022}, or moiré patterns \citep{Mak2022}) and delocalized physics, which leads to emergent states. A key aspect of the phenomenology of correlated electron materials is the interaction between electronic and crystal lattice symmetries \citep{Haule2016,Landscapes,Haule2017Nickelates,Georgescu2019,Olegb, Fowlie2023,structure1,structure2,Subedi2015}. To achieve correlated electron behavior, the level of interactions must be precisely balanced: electron-electron interactions must be comparable to the kinetic energy of the electrons to allow local moments to form emergent states. However, if the states are too isolated, no emergent phenomena occur—similar to a set of isolated atoms in a vacuum. While isolated impurities in semiconductors can behave like color centers (or f-centers, from the German 'Farbe,' meaning color) and provide tunable local properties, their isolated nature makes them better suited for the individual control of each atom rather than for emergent states like superconductivity. Cu$^{2+}$ ions are well-known color centers, often used for pedagogical purposes \citep{Earl1985}.

In the recent case of Cu-doped lead apatite ('LK-99') \citep{KumarAbsenceConditions, wu2023observation,zhu2023order,timokhin2023synthesis,guo2023ferromagnetic,wu2023successful,lee2023superconductor,lee2023roomtemperature,jiang2023pb9cupo46oh2,si2023electronic}, Cu atoms replace 10\% of Pb atoms in an insulating material. A variety of theoretical reports have identified degenerate flatbands associated with the Cu impurity at the Fermi level, in a magnetic, d$^9$ configuration, leading to comparisons with cuprate superconductors. This has sparked further debate about which level of DFT or beyond-DFT calculations is required to obtain the experimentally observed insulating state in this material, as well as whether the material is insulating or not \citep{griffin2023origin, Lai_2023, si2023electronic, tavakol2023minimal, tao2023cu,hao2023firstprinciples,yue2023correlated,si2023pb10xcuxpo46o,HeldPRB,SchoopPRB,Lucian2024,swift2023commentorigincorrelatedisolated}. Several initial theoretical calculations suggest that the experimentally observed doping site is energetically disfavored in DFT, and indicate an over-contraction of the unit cell volume, or that the resulting structure may be dynamically unstable \citep{griffin2023origin,Lucian2024,SchoopPRB}. The results we present in this paper find that a Jahn-Teller distortion stabilizes the experimentally observed doping site, and an insulating ground state.

While we were able to stabilize a metallic state with relatively flat bands at the Fermi level within DFT and DFT+U through Cu doping, with octahedral coordination and trigonal symmetry—similar to previous reports—we find this state to be approximately 0.8 eV higher in energy than the other states discussed in this work, making it energetically disfavored. Since the transition metal states are strongly localized, our results can be interpreted in analogy to the Jahn-Teller theorem, which disfavors partially filled degenerate orbitals \citep{Bersuker2001,Bersuker2}. We note that our DFT results are consistent whether using the PBE exchange-correlation functional with spin magnetization or DFT+U. Unlike other materials where a +U correction or the SCAN exchange-correlation functional may be required to model the transition metal d-sites and resulting band gaps, the tendency for structural symmetry breaking here is so strong that neither is required.

Our results show that a symmetry breaking of the ligand environment around the Cu placed in the Pb site initially reported experimentally, leads to the opening of a gap in the material. This creates an isolated, spin-polarized empty flatband within the band gap, corresponding to the now localized charge. This can be understood as a manifestation of the Jahn-Teller theorem, but rather than occurring in an isolated molecule, it arises from an isolated impurity in an insulator and its immediate ligand environment, leading to a localized charge on the magnetic ion. This structural relaxation reduces the overestimation of the volume contraction, showing that doping either Pb site leads to the same volume contraction. We find that the experimentally observed Cu doping site is slightly energetically favored, suggesting that the resulting material is likely to be an insulating, transparent material with potential Cu$^{2+}$ color centers that may become active at lower temperatures, reconciling the discrepancy between theory and experiment. 

We also perform an example calculation with 20\% doping, placing one Cu atom in each of the two main Pb sites, to investigate whether these trends change. We find that the resulting state continues to be insulating, with localized spin states. This analysis suggests that doping of this kind could lead to non-relativistic spin splitting (often referred to as altermagnetism). Here, the energy separation between spins and their relative isolation may allow for their independent control.

\subsection{Cu-Doping Pb$_{10}$(PO$_4$)$_6$O} 

 \begin{figure}
    \centering
    \includegraphics[width=0.9\textwidth]{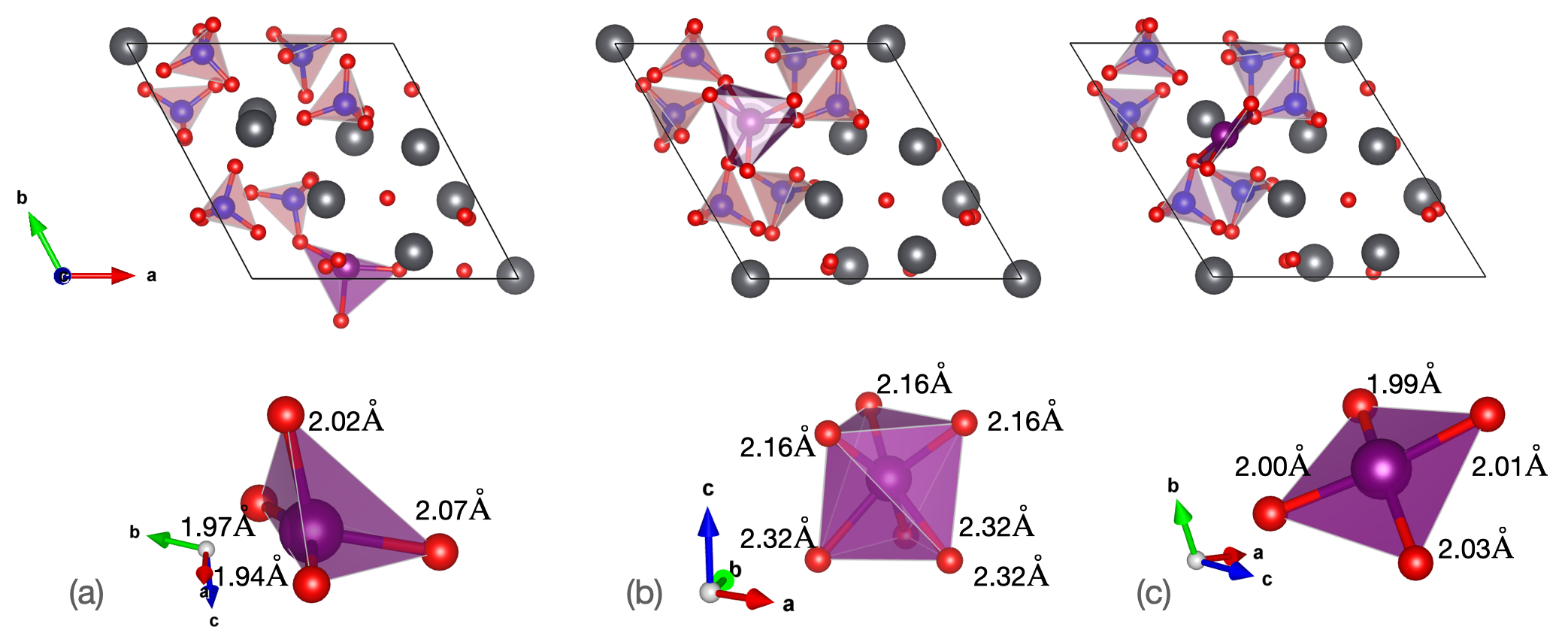}
    \caption{Top: Unit cell of Cu-doped Pb$_{10}$(PO$_4$)$_6$O. Bottom: Cu-ligand structure, with Cu-O bond lengths indicated. (a) Cu replacing a Pb(II) site, resulting in a distorted tetrahedral CuO$_4$ coordination. (b) and (c) show the two possible states of the Pb(1) doping location, namely :(b) High-symmetry structure obtained by replacing Pb(I), consistent with previous work. (c) Lower-symmetry coordination and lower energy after further relaxing the structure in (b), leading to an insulating state. O - red, Cu - purple, P - blue, Pb - dark gray. Cu-O bond lengths are shown for U=0; U=4 eV bonds differ by about 0-0.02$\AA$ from those shown in this figure.}
    \label{fig:structures}
\end{figure}

 \begin{figure}
    \centering
    \includegraphics[width=1.0\textwidth]{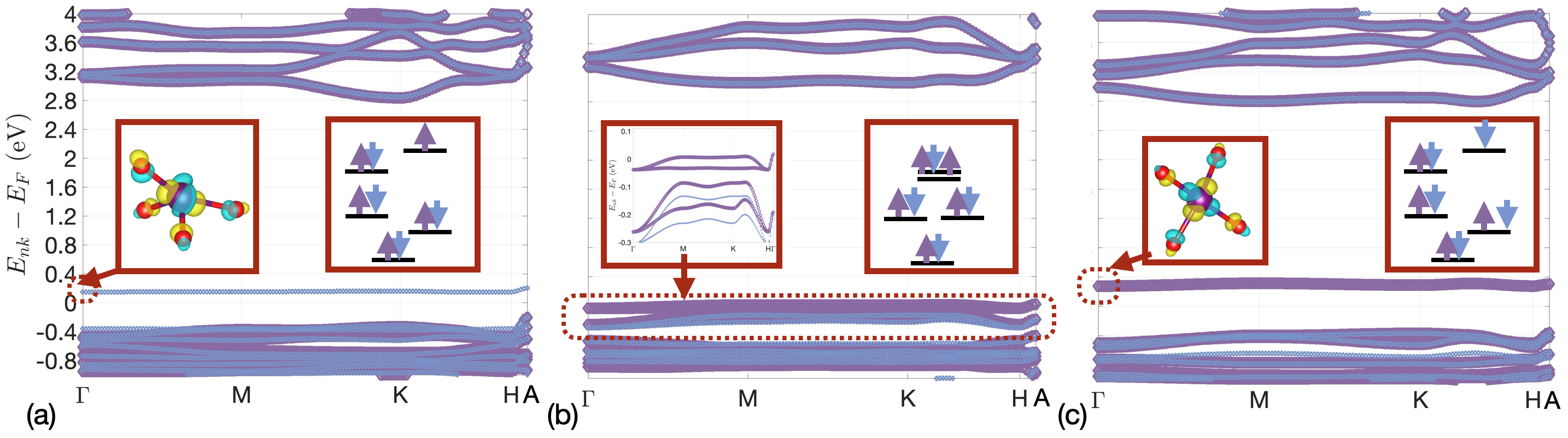}\\
    \caption{(a) Band structure corresponding to Cu replacing the Pb(2) site. The flat band in the gap corresponds to a spin-polarized, localized hole state. Inset: (left) wavefunction plotted in real space at $\Gamma$, with an isovalue of 0.008 showing localization of the spin-polarized hole state on the CuO$_4$ structure, and (right) orbital order and filling of the Cu d orbitals. (b) and (c) refer to the main doping site discussed in this text, namely the Pb(1) location. (b) Bands for the high-symmetry, unstable structure corresponding to Cu replacing the Pb(1) site. The inset (left) shows a zoom-in on the flat bands at the Fermi level, and (right) in the absence of a local Jahn-Teller effect, the material appears to be metallic with two partially filled degenerate orbitals. (c) Bands for the low-symmetry structure with Cu replacing the Pb(1) site, showing a localized hole state on the Cu atom. Inset: wavefunction plotted in real space at $\Gamma$, with an isovalue of 0.008 showing localization of the spin-polarized hole state on the CuO$_4$ structure. Bands are plotted at U=0 eV.}
    \label{fig:bands}
\end{figure}

\begin{table}[]
\begin{tabular}{|l|r|r|r|r|}
\hline
Doping Site & \multicolumn{1}{l|}{U(eV)} & \multicolumn{1}{l|}{Relative Energy (meV)} & \multicolumn{1}{l|}{Volume($\AA^3$)} & \multicolumn{1}{l|}{Contraction $\% $}\\ \hline
No doping     &  N/A &    N/A  & 642.15 & 0\\ \hline
Pb(I) - Low Sym Oct & 0 & 0                        & 634.78  & 0.93 \\ \hline
Pb(I) - High-Sym Oct  & 0 &   558                      & 624.83 & 2.51 \\ \hline
Pb(II) - Tetragonal  & 0 &  127                     & 633.92 &  1.19\\ \hline
Pb(I) - Low Sym Oct & 4 & 0                        & 634.78  & 1.14 \\ \hline
Pb(I) - High-Sym Oct  & 4 &  810                      & 624.83 & 2.70\\ \hline
Pb(II) - Tetragonal  & 4 &  112                      & 633.92 & 1.28\\ \hline
\end{tabular}

\caption{The lowest energy state for each dopant site and symmetry configuration is compared. The fully symmetry-broken Pb(I) site has the lowest energy for both U=4 eV and U=0, consistent with experimental results. In their low-symmetry state, both dopant locations result in a similar volume contraction.}
\end{table}

The undoped lead phosphate apatite structure, as reported on the Materials Project \citep{matproj} Pb$_{10}$(PO$_4$)$_6$O has two symmetry-inequivalent Pb sites, with distinct ligand structures (Figure \ref{fig:structures}). The formula unit, and unit cell, both contain 41 atoms. Doping a site with a Cu atom, and relaxing the structure, will initially lead to either a Cu 6-fold coordinated with O atoms, or a 4-fold coordinated Cu. The site leading to a 6-fold coordinated structure is usually referred to as Pb(1), while the one leading 4-fold coordinated is referred to as Pb(2). 

Doping the Pb(1) site, and its subsequent relaxation from 6-fold to 4-fold ligand coordination is the subject of much of this work, and was the initial site reported experimentally, and studied for its potential flatbands at the Fermi level. By contrast, doping the Pb(2) site was established from early on in the literature as leading to an insulating state.\citep{griffin2023origin}

Regardless of the doping site, the resulting Cu atom is in the Cu$^{+2}$ state, i.e. in a $d^9$ configuration. Recent experimental work has suggested that the doped site is Pb(1), leading to this material being named 'LK-99', with theoretical work suggesting that trigonal crystal field splitting\citep{griffin2023origin,Lai_2023} - is what leads to two narrow bands around the Fermi level (flatbands); while doping the Pb(2) site leads to the opening of a band gap as a result of strong crystal field splitting, with an S=1/2 empty state separated from the 8 Cu d electrons that are in the same energy range as the valence band of the bulk material. The results we obtain by doping the Pb(2) site agree with those previously reported, with a local, low-symmetry Cu environment leading to a strong orbital polarization of the Cu atom, and an insulating state with an S=1/2 localized hole. This can be seen in DFT results as a single flatband, localized on the Cu site, within the material's bandgap. We obtain a total magnetization of exactly $1\mu_B$ integrated over the unit cell. After a Jahn-Teller distortion we describe in detail, we find similar results for the Pb(1) doping site.

When doping the Pb(1) site, we initially find a structure similar to what has been previously reported, exhibiting trigonal symmetry and a metallic phase with relatively flat bands around the Fermi level. These bands correspond to three electrons occupying two degenerate $e_g$ orbitals. The d orbital occupation eigenvalues for the majority spin, as obtained from atomic orbital projectors in the high-symmetry structure, are: 0.996, 0.996, 0.996, 0.999, 0.999. For the minority spin, they are: 0.797, 0.797, 0.994, 0.994, 0.994. This can be interpreted as two filled orbitals and two partially filled orbitals. While orbital projectors are not necessarily orthogonal or normalized and may not provide optimal physical insight, they offer a quick estimation of the system's symmetry. Within standard DFT formalism, this structure has to be metallic.

As discussed by this author and others in the context of 2D transition metal halides, a $d^9$ configuration in a distorted octahedron with trigonal symmetry may exhibit symmetry-lowering instabilities \citep{Alex2D}, likely leading to strong crystal structure symmetry breaking. This would favor a state with a single, half-filled orbital and four fully occupied orbitals.

We can break the symmetry by introducing arbitrarily small, random displacements in the crystal structure to determine whether this state is energetically unfavorable, or by applying a very small force tolerance and removing symmetry enforcement in the DFT calculations. Regardless of approach, we observe an additional symmetry reduction, which opens a band gap, lowers the state's energy, and makes it more energetically favorable than the Pb(2) state (Figure \ref{fig:bands}). This eliminates the apparent flatbands at the Fermi level and leads to an electronic structure similar to that seen when doping the tetrahedral site. As with the Pb(2) site, we obtain a Cu atom coordinated with four oxygen atoms. Another way to interpret this result is as a distorted CuO$_6$ unit, where the Cu-O bond lengths vary from $2 \AA$ to $3.26 \AA$. However we find it more appropriate to refer to this as a CuO$_4$ square, with the other two oxygen atoms further from the Cu. This relaxation lowers the total energy by about 0.8 eV compared to the structure with narrow bands at the Fermi level, making the Pb(1) site slightly favored.
 
From this particular type of calculation, it is unclear whether the driving force of the symmetry breaking is the crystal structure or the electronic structure. Within density functional theory, in the absence of structural symmetry breaking, electronic symmetry breaking is highly unlikely. To fully understand the driving force, calculations beyond band theory, such as dynamical mean field theory (DMFT), which explicitly account for possible electronic symmetry breaking even in the absence of crystal symmetry breaking would be necessary\cite{Landscapes}. In fact, previous DMFT calculations have shown that Cu states are strongly in the Mott regime, even in the high-symmetry Cu environment without further relaxation\cite{Si2023a}. This is expected, given the highly isolated nature of the Cu impurity. Mott insulator models are often applied in the context of isolated charge impurities in solid materials. Nonetheless, within DFT, we find a spin-polarized, localized hole state on the Cu, taking the form of an $x^2-y^2$ orbital, oriented along the Cu-O bonds.

To quantify the electronic symmetry breaking in the two different low-symmetry structures, we can examine the d occupation eigenvalues. Doping the Pb(2) site, i.e., the tetrahedral site, results in spin-up orbital occupation eigenvalues of 0.989  0.996  0.997  0.999  1.000 (note that the occupations are not normalized). The spin-down occupations are 0.431  0.989  0.993  0.995  0.997. For the fully relaxed, low-symmetry lead apatite structure with Cu in the octahedral (or square after full relaxation) site, the spin-up eigenvalues are 0.988  0.998  0.998  1.000  1.002, and the spin-down occupations are  0.437  0.982  0.995  0.996  0.998. These values clearly reflect the broken symmetry at the Cu site, with a single orbital partially filled. Although the ligand field effects should, in principle, differ significantly between the two types of doped sites that lead to an insulating state, the resulting orbital and spin polarization for the two types of dopants is very similar, as is the unit cell volume.

We also compare the unit cell volume contraction between the undoped and doped material, as reported in Table I; doping leads to a smaller contraction than reported in previous work, in better agreement with experiment \citep{griffin2023origin, Lai_2023}. Additionally, we find that the volume contraction is relatively independent of the doping site.

We conclude that doping both the Pb(1) and Pb(2) sites leads to a strongly insulating state, with the Cu $d^9$ impurity state strongly localized in a CuO$_4$ environment. The Cu ion may act as an isolated color center. Our results clearly show a contraction of the unit cell volume, though this contraction does not depend on the doping site. Additionally, we find that the Pb(1) site is slightly favored for doping.

In both cases, the material remains an insulator, with an isolated impurity site and a flat band—similar to what one would expect for an orbital in an isolated atom in vacuum—rather than a metal that could display superconductivity. This is consistent with experimental findings. The minor change in bandgap, or the Cu states, may be responsible for the material's purple color.

 \begin{figure}
    \centering
    \includegraphics[width=0.9\textwidth]{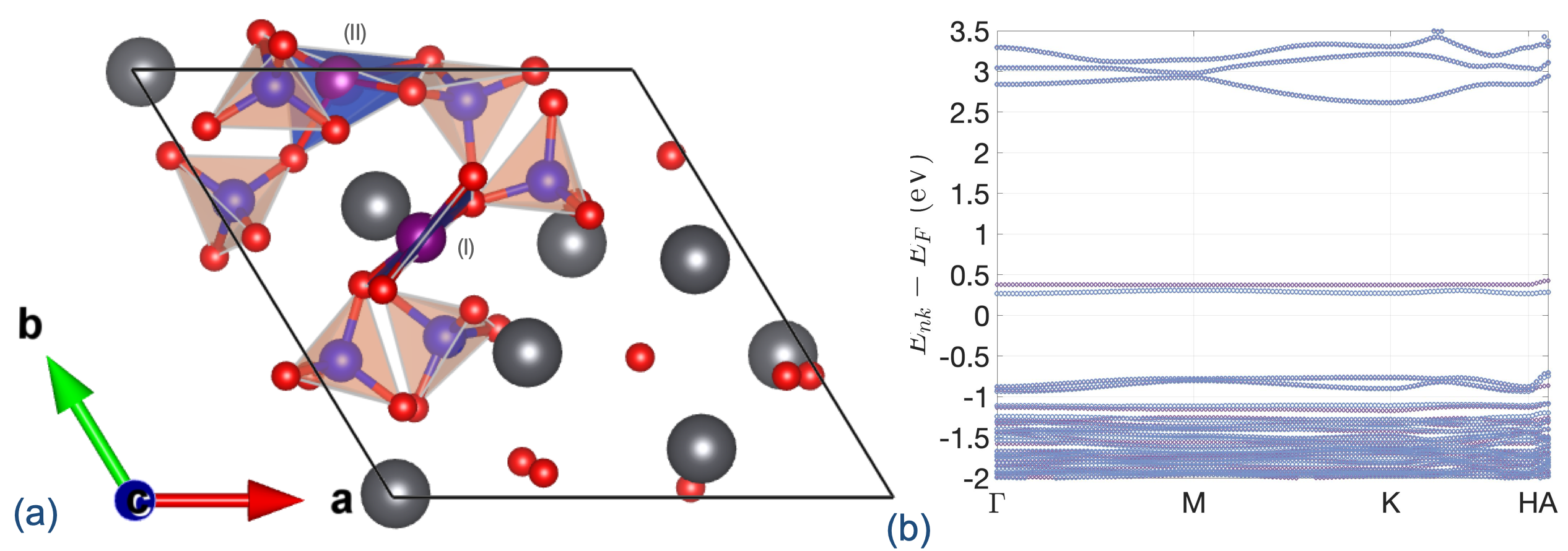}\\
    \caption{(a) Crystal structure with the two types of Cu site dopants marked and (b) band structure after doping by replacing two Pb sites, one of each coordination type, with opposing spins in lead phosphate apatite. The difference in ligand coordination between the two Cu atoms results in non-relativistic spin splitting between the two impurity hole states.}
    \label{fig:2imp}
\end{figure}

The possibility of doping sites with different symmetries makes this material an intriguing example of non-relativistic spin splitting (or altermagnetism) \citep{Rashba2020, Yuan2023, bhowal2022magnetic, Altermagnetism, Mazin, hariki2023xray}, though in a different context of isolated charges. Specifically, it refers to spin splitting in an antiferromagnetic state, in the absence of spin-orbit coupling, as a result of geometry. Altermagnetism is a growing field, and has recently been expanded to also expanded to include non-collinear spins \citep{addaltermagnetism}, with additional discussions of collinear non-relativistic spin splitting that also leads to band splitting at the $\Gamma$ point in other classes of materials discussed in other recent work.\citep{Yuan2024}

While both Cu sites are 4-fold coordinated, one is tetrahedral and the other is square. This leads to different crystal field splitting, and the two states exhibit a small energy separation along the band path, suggesting they could be controlled independently. We show this in Figure \ref{fig:2imp}.

We also observe that, even at a 20\% concentration, the impurities behave as two localized charge carriers, maintaining the material's insulating properties.

\section{Conclusions and Discussion}
Key to the design of correlated electron materials is achieving the 'just right' balance between local electron-electron interactions and the kinetic energy of the electrons. While flatbands in various materials are gaining popularity, bands that are too narrow correspond to isolated states—similar to the states of an atom in vacuum or an isolated impurity in an insulator within the Mott picture. 

Since the discovery of superconductivity in copper oxides, based on the relatively localized bands of the Cu $x^2-y^2$ orbital, the search for similar electronic mechanisms has led to the search for superconductivity in flatbands materials – for example in multilayer Moire graphene.\citep{moiregraphenescience,moiregraphenereview,moirebands,Cao2018}

In superconducting cuprates, for example, the Cu d are relatively close – only separated by an oxygen ligand. The bands are relatively broad ($~$1eV), however they are comparable with the local electron-electron interactions.

In materials with topological flatbands arising from frustrated hopping, for example Lieb and Kagome lattices\citep{LiebAndKagome,StrainLieb,KagomeFrustration,KagomeWilson,KagomeStrain}, the relevant electronic states overlap, and emergent states are possible. The atomic flat bands on the Cu ion in this material are orbitals localized on the atom, with the Cu atoms separated by 7-9 $\AA$ (the length of the unit cell lattice vectors). This does not allow for the formation of emergent correlated states from the local quantum states on the Cu.

If localized electronic states exhibit degeneracy at the Fermi level, an electronic and ionic symmetry lowering is likely to occur, as a result of the Jahn-Teller theorem, which removes the degeneracy and further localizes the electronic states. Focusing on the example of LK-99, we find that while the localized charge with an S=1/2 spin may be technologically relevant, it is unlikely to result in metallic—let alone superconducting—behavior. As a result, 'just right' in the case of doped materials may also include the amount of charge added via transition metal sites. On the other side, however, these isolated dopants may be useful as possible spin and color centers. 
DFT+DMFT calculations may further help elucidate the electron-crystal structure coupling in this class of materials\citep{DFTDMFTSpectral}, as well as provide detailed understanding of their spectral function, and detailed many-electron behavior, versus interaction strength; DFT+DMFT calculations may also allow for a study of the role of electronic entropy in possible phase transitions in this material versus temperature. \citep{temperaturecrystal,Landscapes}

\section{Methods}
Our calculations are performed using the Quantum Espresso \citep{QE} version 6.8 code, on the Quartz and Big Red 200 clusters hosted at Indiana University. We employ ultrasoft pseudopotentials \citep{ultrasoft}, a 4x4x4 k-mesh, and an energy cutoff of 544 eV. The DFT+U method is used in the Dudarev approximation, with forces relaxed to $10^{-4}$ eV/Å, and the default 'atomic' projectors are utilized to enable fast structural optimization and prevent numerical instabilities. We apply a U value of 0 eV for most calculations in this work, unless stated otherwise, when calculations are done with U=4eV to compare with previous work. However, we note that given the extremely localized nature of the Cu state, which is found in a wide bandgap material, this U value may be underestimated. 

All calculations performed in this paper are performed by allowing spin polarization. This allows the Jahn-Teller mode to break symmetry, and lead to the distorted structure discussed in this manuscript.
Although our calculations focus on a doped variant of lead apatite, the hydroxy form $Pb_{10}(PO_4)_6(OH)_2$ is more stable in air and thus more representative of real samples. However, the key features—the isolated Cu site substituting Pb, the Pb site symmetry, and the Cu valence state—remain unchanged, so our conclusions still apply. Previous theoretical and experimental investigations\citep{SchoopPRB} have reported the same electronic structure around the Fermi level for the high symmetry structure and the Pb(2) doped location site, further supporting the relevance of our findings.
In this manuscript, we use “vanilla” spin-polarized DFT—without additional corrections or advanced functionals (e.g., +U, HSE, SCAN)—to demonstrate the strong tendency of this material to adopt a lower-symmetry electronic and crystal structure around the Cu dopant. Although we employ spin polarization within DFT to open an insulating gap and achieve a symmetry-broken state, this does not imply actual magnetic order. Rather, the disordered nature of the doped material and the localized charge make an emergent magnetic state unlikely.\\
It is true that DFT+DMFT can predict an insulating state in the absence of structural distortion through Mott mechanisms; however, recent work shows that DFT+DMFT and DFT+U yield similar orbital polarization and structural distortions, as DFT+U reproduces DFT+DMFT in the Hartree Fock limit  \citep{carta2024explicitdemonstrationequivalencedftu}. Even in DFT+DMFT studies, correctly modeling the ground state requires accounting for structural relaxation and electron-lattice coupling.\citep{Landscapes,temperaturecrystal }.
Although DFT+DMFT could provide further insight into the spectral function, our standard DFT calculations are sufficient for capturing the key physics here—namely, the electron- and crystal-structure symmetry breaking and the associated charge localization.

Example crystal structures can be found at \href{https://github.com/alexandrub53/DopedInsulators}{Github}.

\section{Acknowledgments}
We thank Jennifer Fowlie and Divine Kumah for their discussions during the preparation of this work, as well as Leslie Schoop and Andrew McCalip for discussions on the most recent experimental status of the Cu-doped lead phosphate apatite material. We also thank Markus Suta for valuable discussions on the interplay of crystal-lattice coupling and temperature limits on the luminescence of color centers in this type of material.

This research was supported in part by Lilly Endowment, Inc., through its support for the Indiana University Pervasive Technology Institute. The computations presented in this work were performed on the Big Red 200 and Quartz supercomputer clusters at Indiana University.

\begin{tocentry}
\includegraphics[height=1.15in]{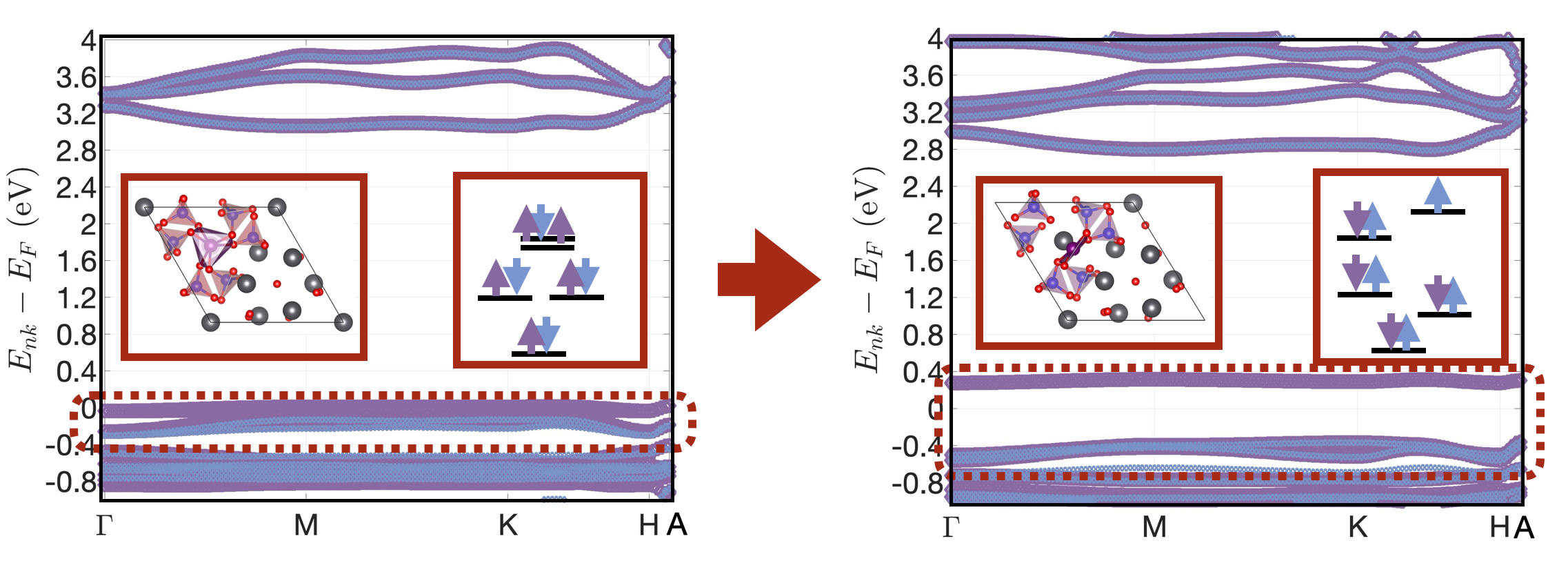}
\end{tocentry}
\bibliography{Mendeley2,more_refs2}

\end{document}